\documentclass[twocolumn,floatfix,superscriptaddress,amsmath,showpacs,showkeys,aps,prb]{revtex4}

\usepackage[normalem]{ulem}
\usepackage{t1enc}
\usepackage[final]{graphicx}
\usepackage{graphicx}
\usepackage{epsfig}
\usepackage{bm}
\usepackage{color}
\usepackage{hyperref}

\newcommand{\avrg}[1]{\left\langle #1 \right\rangle}

\begin{document}

\title{Smart random walkers: the cost of knowing the path}

\author{Juan I. Perotti}
\email{perotti@famaf.unc.edu.ar}
\affiliation{Facultad de Matem\'atica, Astronom\'{\i}a y
F\'{\i}sica, Universidad Nacional de C\'ordoba and \\ Instituto de
F\'{\i}sica Enrique Gaviola  (IFEG-CONICET), Ciudad Universitaria,
5000 C\'ordoba, Argentina}

\author{Orlando V. Billoni}
\email{billoni@famaf.unc.edu.ar}
\affiliation{Facultad de Matem\'atica, Astronom\'{\i}a y
F\'{\i}sica, Universidad Nacional de C\'ordoba and \\ Instituto de
F\'{\i}sica Enrique Gaviola  (IFEG-CONICET), Ciudad Universitaria,
5000 C\'ordoba, Argentina}

\date{\today}

\begin{abstract}

In this work we study the problem of targeting signals in networks using  
entropy information measurements to quantify the cost of targeting. We introduce 
a penalization rule that imposes a restriction on the long paths and therefore focus
the signal to the target. By this scheme we go continuously from fully random walkers to 
walkers biased to the target. We found that the optimal degree of penalization is
mainly determined by the topology of the network.  By analyzing several examples,
we have found that a small amount of penalization reduces considerably the typical walk length,
and from this we conclude that a network can be efficiently navigated with restricted information.   

\end{abstract}


\maketitle

\section{Introduction}
\label{intro}


In the problem of targeted signaling or targeted navigability in a network,
a message or vehicle begins a journey at a given source vertex with the intention
of reaching another target vertex in the most efficient way possible. It is
implicitly  assumed that the message or vehicle is restricted to jumping from vertex 
to vertex along the edges available in the network.
%
The applications of this area, such as 
distant communication in complex systems \cite{Sneppen2005}, 
and problems related to the traffic in cities \cite{Rosvall2005b,Scellato2010},
make the area an active field of research.
%
The efficiency in solving the problem is measured in terms of a cost, which is associated 
with each possible path the message or vehicle can follow in its journey from the source to the target.
Two main issues have to be accounted for in defining the cost of targeting:
the length of the paths and the difficulty of identifying a set of convenient paths which connect the 
source with the target.
Once the cost is defined, the efficiency in the task of targeted signaling or navigability can be 
improved by choosing a convenient searching strategy that minimize the cost.

The most basic strategy for searching the target is the non-biased or fully random walker.
The message or vehicle moves randomly without bias through the network
with the hope that eventually it will reach the target.
In this case, there is no cost associated in choosing the appropriate path to
the target, so the focus is in the determination of how long it will
take to the message or vehicle to reach its target. It means that the efficiency is 
determined by comunicability between the source and the target. This problem was studied in detail
by Estrada et al. \cite{Estrada2008, Estrada2012}, who introduced a penalty based in the lenghts of the paths.   
Even more, this is a problem related to the first passage time
and has been investigated in several paradigmatic network models \cite{Rieger2004}.
%
%
It is important to stress here that the random walk strategy
has to be differentiated from network sampling using random walkers\cite{Costa2007}
and from non-specific broadcasting where a signal is propagated and amplified
as in the spreading of diseases, spam or computer viruses \cite{Moreno2003,Balthrop2004}.

Other approaches deviate from a fully random walker but still dismiss the quantification
of the cost of choosing the appropriated paths from the source to the target.
%
Now the problem is finding a reasonable strategy for searching the target,
and the selection of a particular strategy is driven by minimizing of the length of the journey 
from the source to the target. Among the different strategies are self--avoiding random walks\cite{Yang2005},
intermittent random walks on lattices\cite{Oshanin2009}, the consideration of local topological features\cite{Adamic2001},
and greedy strategies, which are used in the case of networks with spatial embedding\cite{Kleinberg2000,Moura2003}.
A common feature of all the above mentioned methods is that they use knowledge about the topological structure 
for the design of the strategies.

A different approach to the problem consists in evaluating the difficulty of choosing the appropriated
paths. The difficulty can be quantified in terms of the amount of information or knowledge
required to follow these paths.
%
In some cases the information is measured ad hoc, for example, using a fixed information
cost per vertex traversed \cite{Cajueiro2009,Cajueiro2010}.
In general, the information required for choosing the right direction in the network
depends on the local topological details; for instance, the information required to
take the right direction grows with the number of available options.
%
Entropy measures provide a natural way of quantifying information;
in fact, this  measurements have been applied successfully in complex 
networks before \cite{Sole2004,Demetrius2005,Bianconi2009,Lin2010,Enrico2011}. 
%
In particular, these methods were applied recently to quantify information in the problem of targeted
signaling or navigability \cite{Rosvall2005,Rosvall2005a,Rosvall2005b} and in the complementary problem of efficient 
diffusion in a network \cite{Burda2009,Sinatra2011}.
This is the approach  we adopted in our work to quantify the difficulty of choosing 
the appropriate paths that connect the source with the target.
We focus on strategies that may be adapted
to any topology and in which the message or vehicle is represented by a biased
random walker. These strategies will allow the interpolation between a fully random walker,
which uses no information to reach its destiny, and a directed walker that travels
along the shortest paths using all the available information to orient itself.
In this regard, there are several antecedents with strategies that interpolate
to some extend the random and the biased regime \cite{Sood2007, Fronczak2009, Cajueiro2009, Rosvall2005a,Rosvall2005,Rosvall2005b}. 

In this work we follow the line of previous works \cite{Cajueiro2009,Rosvall2005a} in the sense
that an information measure is used to regulate how directed the walks are.
%
%
We extend the ideas of Refs. [\onlinecite{Rosvall2005,Rosvall2005b}] where the information
is measured considering only the shortest paths by allowing the usage of less information at expenses
of longer walks \cite{Rosvall2005a}.
%
We introduce a formalism for measuring the amount of information
used by a biased random walker to reach its target. 
Using this formalism we develop a method which depends on one parameter 
that regulates how biased is the random walk. 
In our method the overall information is increased  each time the walker performs a step, 
so longer walks result in larger penalization.
Optimizing the walker's information forces the walker to travel along increasingly shorter paths
or, equivalently, the paths are biased to the target.

The paper is organized as follow. In section \ref{theo_back}
we introduce the theoretical background, defining the  
measures of information used by the random walker in going from the source to the target.
We also introduce the penalization rules used to interpolate between the random and the directed
regimes; in particular, the optimal penalization is defined.
In section \ref{analytical} we analyze simple examples that can be solved analytically, which are 
useful to understand how the method works in different topological environments, including some
limiting cases.
In section \ref{simulations} we applied the method by using numerical simulations in more complex 
networks, such as a random network and a Barab\'asi--Albert scale--free network model. 
Finally, in section \ref{conclu} final remarks, conclusions
and possible extensions to our work are discussed.

\section{The Model}
\label{theo_back}

Consider a non-directed network with $N$ vertices and $M$ links where a random 
walker jumps at a given time step from a vertex $i$ to a neighbor vertex $j$ with probability $q_{ij}$. 
For each vertex $i$  in the network the transition probabilities $q_{ij}$ satisfy 
the normalization condition,
\begin{equation}
\label{eq1}
\sum_{j\in nn_i} q_{ij} = 1,
\end{equation}
\noindent where $nn_i$ is the set of all nearest-neighbors vertices of vertex $i$, 
and $q_{ii}=0$ for  all $i$. This means the walker is forced to move at each time step. 
The amount of information given to the walker for taking an exit from a given vertex $i$ 
to one of its nearest neighbors is the information cost defined by \cite{Rosvall2005}
\begin{equation}
\label{eq2}
\ln(k_i) - \left[ - \sum_{j\in nn_i} q_{ij}\ln q_{ij} \right], 
\end{equation}
which is the difference between the maximum entropy in the space of events of taking one of the exits
minus the entropy the exits of vertex $i$ already have associated. Here $k_i$ denotes the degree of vertex $i$. 
Let us consider now that the walker starts its journey at a source vertex named $s$ and ends the trip at a vertex  
we call the target $t$; furthermore, during the journey the walker passes by the vertex $i$. We want to
obtain an expression for the information needed in going from $s$ to $t$ given a distribution of
probabilities $q_{ij}$. 
From the information cost defined above one can derive a recursive expression for the amount 
of information $S(i\to t)$ used by the walker in going from vertex $i$ to $t$. Accordingly this 
information cost is expressed as
\begin{equation}
\label{eq3}
S(i\to t) = \ln(k_i) + \sum_{j\in nn_i} q_{ij}\ln q_{ij} + \sum_{j\in nn_i} q_{ij}S(j\to t).
\end{equation}
\noindent Hence, with the constrain that $S(t\to t) = 0$ (i.e., no information is needed by the
walker once the target is reached), a set of linear equations  
with unknowns $\{ S(i\to t)\}_{i=1,...,N}$ can be defined 
and solved provided the probabilities $\{q_{ij}\}$ are known. A similar approach was 
used by Rosvall et al. [\onlinecite{Rosvall2005}] to quantify  the amount of information needed by a walker 
which is restricted to walking only the shortest paths. In the case
that the random walker can step back during the walk, the amount of information is\cite{Rosvall2005a}
\begin{equation}
\label{eq4}
S_{sp}(s\to t) = -\ln\left( \sum_{\pi \in \Pi(s,t)} \frac{1}{k_s}\prod_{j\in \hat{\pi}} \frac{1}{k_j}\right),
\end{equation}
\noindent where $\Pi(s,t)$ denotes the set of all shortest paths $\pi$ between $s$ and $t$,
and $\hat{\pi}$ denotes the set of interior vertices of the shortest path $\pi$.

The minimum for the information $S(s\to t)$ introduced in Eq. (\ref{eq3}), 
regarding the transition probabilities $q_{ij}$, corresponds to a fully random
walker with probabilities defined by
\begin{equation}
\label{eq5}
q_{ij} = \frac{1}{k_i}, \;\;\; \forall i,j.
\end{equation}
\noindent In this case $S(i\to t) = 0$ for all $i$. As expected, in finite networks the
fully random walker needs no information to reach the target, but this has the drawback 
of leading to very long walks on average. Since we are interested in targeted signaling, the results
obtained above are of little utility. In order to fix this problem, we introduce 
a penalization rule that weights the paths favoring the shortest 
paths to the target. This penalization will modify the transition probabilities that minimize
the information required to reach the target; the longer walks will be rejected and then
a random  walker that searches the network using this probabilities will be biased to the target. 

The simplest way to introduce a penalty is by paying a cost each time the walker 
passes through a vertex. This information cost is not used by the walker when it is travelling 
the network--unlike the information associated to the $q_{ij}$--but it allows the evaluation 
of intrinsic properties of the paths to the target, taking into account the whole 
network. For instance, depending on the degree of penalty needed for reaching an optimal set of
paths, one can estimate the difficulty of finding the paths in a given network.   
Once the penalization term is  introduced in equation (\ref{eq3}), it becomes
\begin{equation}
\label{eq6}
F_{\gamma}(i\to t) = \ln \gamma + \ln(k_i) + \sum_{j\in nn_i} q_{ij}\ln q_{ij} + \sum_{j\in nn_i} q_{ij}F_{\gamma}(j\to t),
\end{equation}
where the term, $\ln \gamma$, with $\gamma \ge 1$ is the penalization term.
Now since $\gamma > 1$,  $F(s\to t) = 0$ is not a minimum anymore and hence a fully random
walk does not minimize the information. As shown in the next section, minimizing $F(s\to t)$ with respect to 
$\{q_{ij}\}$ keeping $\gamma$ fixed leads to a biased walk, which becomes more 
directed to its target as $\gamma$ increases. In fact, the fully
random walker corresponds to $\gamma=1$, and in the other extreme when $\gamma \to \infty$  
the walker is forced to walk along the shortest paths.
The quantity $F_{\gamma}(s\to t)$ stands for the amount of information the
walker uses in going from $s$ to $t$ plus the {\em intrinsic} information related to 
the penalization. 

To clarify the role of $\gamma$ let us introduce a quantity that will allow us to define
an optimal value for the penalization. First of all, we name by 
$\{q^*_{ij}\}$ the probabilities $\{q_{ij}\}$ that minimizes
$F_{\gamma}(s\to t)$ at a given fixed value of $\gamma$, and $F^*_{\gamma}(s\to t)$ 
is the function evaluated at these values, that is, the minimum. Furthermore,  $S^*(s\to t)$ is the value of
$S(s\to t)$ evaluated on $\{q^*_{ij}\}$. We compute the amount of information
introduced by $\gamma$  as $F^*_{\gamma}(s\to t) - S^*(s\to t)$ which is related
to intrinsic properties of the network, as we mentioned above.
Then the relative amount of intrinsic information in going from $s$ to $t$ is
\begin{equation}
\label{eq7}
R_{\gamma}(s \to t) = \frac{F^*_{\gamma}(s\to t)-S^*(s\to t)}{F^*_{\gamma}(s\to t)}.
\end{equation}
The quantity $R_{\gamma}(s\to t)$ lies in $(0,1]$ reaching its maximum value $1$ when
$\gamma \to 1$ or $\gamma \to \infty$. It has a minimum value $R^*(s\to t)$ at $\gamma^*\in (1,\infty)$,
which we define as the optimal value of $\gamma$. At $\gamma^*$ the walker minimizes 
the relative amount of intrinsic information with respect to the whole information. It means that
up to minimum point, the information the walker gains above the paths to the target is preponderating. 
An increase of $\gamma$ further $\gamma^*$ certainly implies a gain of useful information, but at lower
pace than the intrinsic information. Then, the value $\gamma^*$ gives insights about
the searchability of a network in relation to its topology.

\section{Simple examples solved analytically}
\label{analytical}

\begin{figure}
\epsffile{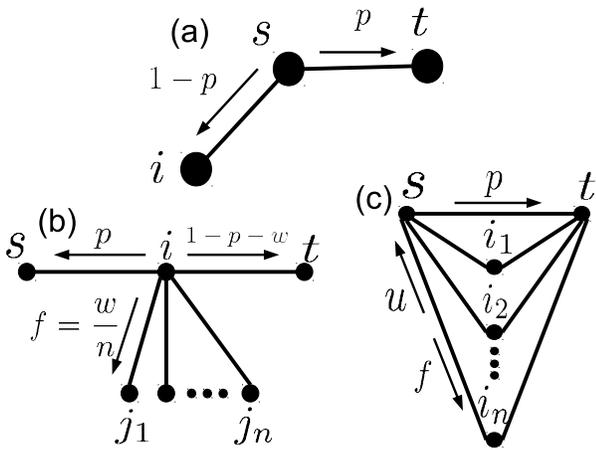}
\caption{\label{fig1}   Simple targeted walks in different network's environments
which can be solved analytically.  Unique path {\bf (a)}, star web {\bf (b)}
and equivalent paths {\bf (c)} (here $nf=1-p$).}
\end{figure}

To further clarify the formal ideas introduced in the above section, 
let us analyze  some simple examples which can be solved analytically. 
We named  each example analyzed in order to facilitate the discussion (see Fig. \ref{fig1}).
In addition, these examples will provide some insights on the problem of targeted
delivery of information or navigation. In particular, the last two examples 
correspond to extreme cases in which remarkably different topological patterns prevail. 
On one extreme is the case where only one right path to the target exists (all 
the other alternative paths dead end), and on the other extreme is the case where there 
are a lot of similar paths to the target--not all of them  optimal--with a few shortest paths. 
The penalization scheme behaves 
differently in each case, serving as an indicator of which kind of topological 
pattern could prevail on real networks or network models.
%
%
\subsection{The unique path}
\begin{figure}
\epsffile{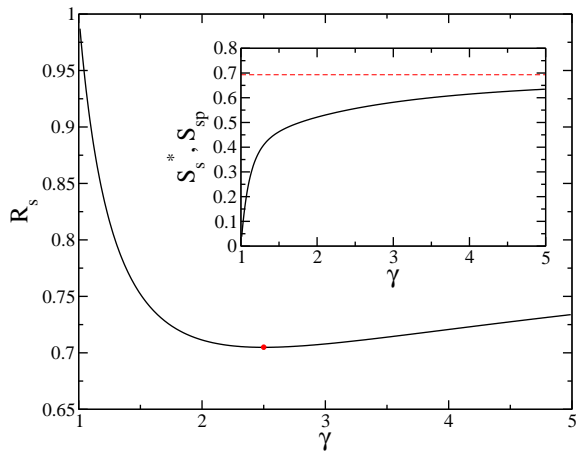}
\caption{\label{fig2}
(Color online) Relative amount of intrinsic information $R_s$ vs. $\gamma$ for the
{\it unique path} [see Fig. \ref{fig1}(a)].
The curve reaches a minimum  value ($R^*_s \simeq 0.7$)
at $\gamma^* \simeq 2.5$ (indicated with a dot) which corresponds to the optimal degree of
penalization. {\em Inset:} Optimal information $S^*_s$ (full line)
and shortest path information $S_{sp}$ (horizontal red dotted line).}
\end{figure}

This example is outlined in Fig. \ref{fig1}({\bf a}). For the sake of 
clarity let us simplify the notation redefining $F_{\gamma}(i\to t)$ by $F_i$, 
$S(i\to t)$ by $S_i$, $R_{\gamma}(i\to t)$ by $R_i$.
In this case the set of equations (\ref{eq6}) takes the form:
\begin{eqnarray}
\label{eq8}
F_s & = & \ln\gamma + \ln 2 + p\ln p + (1-p)\ln(1-p) + (1-p)F_i, \nonumber \\
F_i & = & \ln\gamma + F_s.
\end{eqnarray}
By solving the equation for $F_s$ and minimizing  with 
respect to $p$ fixing $\gamma$,  one obtains the following 
expression for the critical $p$
\begin{equation}
\label{eq9}
p^* = \frac{1}{2}\frac{2\gamma^2-1}{\gamma^2}.
\end{equation}
It is easy to verify that $p^*\to 1/2$ when $\gamma \to 1$ and that $p^* \to 1$ 
when $\gamma\to \infty$; therefore one obtains the expected limiting cases. 
The unpenalized case, $\gamma \to 1$, corresponds to a fully random walker, and the other case 
$\gamma \to \infty$, corresponds to a walker fully biased towards the 
shortest path. One can see that $p^*$ increases  when $\gamma$ goes from $1$ to $\infty$,
indicating that the bias in the walk grows with the penalization $\gamma$; in other words,
a larger penalization leads to a shorter walk. 
The relative amount of intrinsic information corresponding to this example, 
\begin{equation}
\label{eq10}
R_s = \frac{(2\gamma^2+1)\ln\gamma}{(2\gamma^2+1)\ln\gamma - \ln\gamma^2 + (2\gamma^2 - 1)\ln\left(\frac{2\gamma^2-1}{\gamma^2}\right)},
\end{equation}
\noindent is plotted as a function of $\gamma$ in Fig. \ref{fig2}. 
One can see a minimum which corresponds to an optimal penalization $\gamma^*$. 
Notice that $R$ is large even at $\gamma^*$; more than half of the total information $F^*$
is due to the information introduced by $\gamma$. The inset shows the information used by
the walker to reach the target $S^*$ as function of $\gamma$. It increases as $\gamma$
increases and converges asymptotically to the case of the shortest path  $S_{sp} = \ln 2$ as $\gamma \to \infty$.
%

\subsection{The star web}
\begin{figure}
\epsffile{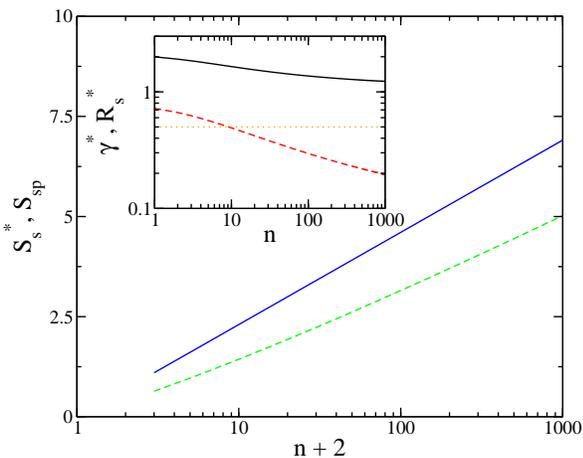}
\caption{\label{fig3}
(Color online) Information as function of the network's size $n$ in a linear-log plot, corresponding
to the {\it star web} [see Fig. \ref{fig1}(b)].
Blue full line shortest path information $S_{sp}$, and the green dashed line is the optimal information $S^*_s$.
{\em Inset:}
The optimal penalization $\gamma^*$ (full black line).
Relative amount of intrinsic information $R^*_s$ (red dashed line).
Horizontal orange dotted line is the reference value for $R^*_s$ at $0.5$.
}
\end{figure}

The starlike network [see Fig. \ref{fig1}(b)] represents the extremal 
case of only one direct path to the target and a large number of dead ends. 
At variance with the previous examples, which contain a fixed number of vertices, this 
example has no restrictions in the number of vertices, allowing  
the study of  quantities that scales with the network's size. 
Like the previous example, this new one can be  analytically solved by
using the particular symmetries of this network, whatever the size of the network.
Equations (\ref{eq6}) in this case reduce to:
\begin{eqnarray}
\label{eq13}
F_i & = & \ln\gamma + \ln(n+2) + p\ln p + (1-p-w)\ln(1-p-w), \nonumber \\
& & + w\ln\frac{w}{n} + pF_s + wF_j, \nonumber \\
F_j & = & \ln\gamma + F_i, \nonumber \\
F_s & = & \ln\gamma + F_i,
\end{eqnarray}
\noindent where $F_j:=F_{j_1}=...=F_{j_n}$. The critical probability 
of $F_s$ is
\begin{equation}
\label{eq14}
p^* = \frac{1}{(n+2)\gamma^2} \;\;\; \mbox{and} \;\;\; w^*=\frac{n}{(n+2)\gamma^2},
\end{equation}

\noindent which satisfies the expected  limiting cases: $p^*\to 1/(n+2)$, $w^*\to n/(n+2)$
for $\gamma\to 1$ and $p^*,w^*\to 0$ for $\gamma \to \infty$. In this example, it is  
interesting to compute the amount of  information related to shortest paths  $S_{sp}$ 
and the optimal information $S^*$ at $\gamma^*$ as function of the number of vertices $n$.
The  following  expression is obtained from the first,  $S_{sp}(s\to t) = \ln(n+2)$, while the second ($S^*$) 
is obtained numerically.   Figure \ref{fig3}  shows these quantities as function of $(n+2)$ in a linear log plot. 
One can see that also $S^*$ scales logarithmically with the network's size; however,
this amount of information is always 
smaller than the information related to the shortest path and the difference between them increases with $n$. 
This implies that, as far as the optimal walks defined by $\{q_{ij}^*\}$ are convenient, it is useful to relax
the restriction of walking the shortest paths, because it is cheaper in terms of information to walk along paths 
which are not so short.
Furthermore, the relative amount  of intrinsic information $R^*_s$ decreases with
the system size (see the inset of Fig. \ref{fig3}). It means that the 
walker's information about the shorter paths in the network eventually becomes predominant. 
This is also consistent with the decrease in the value of optimal penalization 
$\gamma^*$ as function of $n$,  which is required to learn the shorter paths (see the inset).
This indicates that in this topology the walker can learn efficiently the ways to the
target.
%
%
\subsection{The equivalent paths}
\begin{figure}
\epsffile{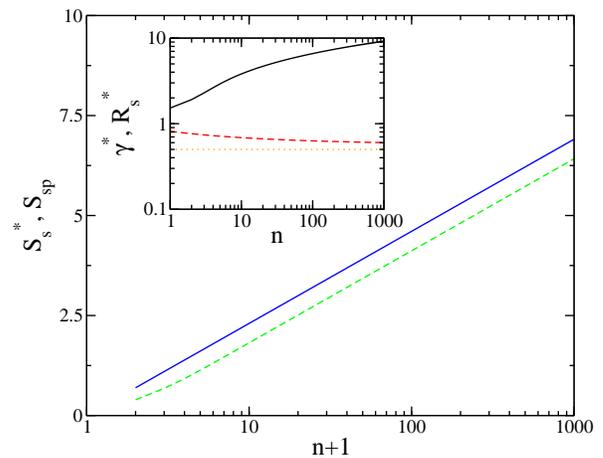}
\caption{\label{fig4}
(Color online) The optimal information $S^*_s$ as a function of system size $n$ (green dashed line),
corresponding to the  {\it equivalent paths} [see Fig. \ref{fig1}(c)].
This information is compared to the shortest path information $S_{sp}$ (blue full line).
Both scale logarithmically with $n$ as indicated by the straight lines in the linear-log
plot, and the difference  $S_{sp}-S^*_s>0$ remains constant.
{\em Inset:} The optimal penalization $\gamma^*$ (black full line) grows
with $n$. The relative amount of intrinsic information predominates
$R^*_s > 0.5$ for all the sizes (red dashed line). The orange dotted line indicates $R^*_s = 0.5$.
}
\end{figure}

This example [see Fig. \ref{fig1}(c)] allows to visualize one of the main 
motivations to generalize the approach which measures the information considering only
shortest paths $S_{sp}$\cite{Rosvall2005,Rosvall2005a} to a measure that
includes all the possible paths. 
Specifically, this example allows the study of a case in which the walker has 
many alternatives consisting of equivalent paths that are not much longer than the shortest ones. 
Due to its particular topology, this example can also be solved analytically
for arbitrary network sizes. Accordingly, applying Eq. (\ref{eq6}) to this particular example,
the following set of equations is generated:
\begin{eqnarray}
\label{eq15}
F_s & = & \ln\gamma + \ln(n+1) + p\ln p  \nonumber \\
& & + (1-p)\ln\frac{(1-p)}{n} + (1-p)F_i \nonumber \\
F_i & = & \ln\gamma + \ln 2 + u\ln u \nonumber \\
& & + (1-u)\ln(1-u) + uF_s,
\end{eqnarray}
\noindent where $F_i:=F_{i_1}=...=F_{i_n}$. As usual we solve for $F_s$ and
minimize, leading to
\begin{equation}
\label{eq16}
p^* = \frac{2\gamma^2(n+1)-n}{(\gamma(n+1)+1)\left(n+\frac{2\gamma^2(n+1)-n}{\gamma(n+1)+1}\right)} 
\end{equation}
\noindent and
\begin{equation}
\label{eq17}
u^* = 1 - \frac{1}{2}\frac{2\gamma^2(n+1)-n}{\gamma(\gamma(n+1)+1)}
\end{equation}
\noindent which satisfies the right limits $p^*\to 1/(n+1)$, $u^*\to 1/2$ for
$\gamma \to 1$ and $p^*\to 1$ ,$u^*\to 0$ for $\gamma\to \infty$. Similarly to the {\it star web},
in this  example the optimal information 
$S^*$
and the shortest paths information
$S_{sp}$ scale logarithmically with the network's size (see Fig. \ref{fig4}), and
also $S_{sp}>S^*$ but at variance with {\it star web} the difference between 
them remains almost constant. 
Here, the optimal penalization $\gamma^*$ grows with system size $n$ and the 
relative amount of intrinsic information $R^*$ is always predominant (see the inset of 
Fig. \ref{fig4}).
These results confirm the intuitive insight that in this kind of topology it is difficult for
the walker to learn the optimal walk pattern, which is related to the fact that discrimination 
between several similar alternatives is expensive.
A comparison between this example and the previous one reveals other important differences in connection
with their topologies. When dead ends prevail in a network, the optimal 
paths are easily achievable in an efficient way; a small penalty is enough and the relative 
amount of intrinsic information is not predominant. On the other hand, if the alternative paths 
prevail, then the optimization procedure is inefficient; a large penalty is required and 
the amount of intrinsic information is predominant. 
%

\section{Numerical simulations results}
\label{simulations}
In this section we apply  the ideas introduced in the previous sections
to more complex network topologies. We choose two paradigmatic cases,
especifically, the random and  scale--free networks. Since these systems cannot be solved
analytically, all the results we show here are obtained by numerical simulations. 
We performed an optimization procedure minimizing $F_s$ with
respect to $\{q_{ij}\}$ for a sequence $\gamma-1 = \delta,2\delta,3\delta,...$
where $\delta$ is a small quantity ($\delta \in [0.005,0.05]$). This is
a convenient procedure since $\{q_{ij}^*\}$ varies smoothly with $\gamma$. We start by
using the values given by equation (\ref{eq5}) as the initial guess for $\gamma = 1+\delta$, 
and then we use the last minimum obtained for the subsequent values of $\gamma$. 
We perform the minimization using the implementation of the SLSQP \cite{Kraft1988,Kraft1994} 
algorithm provided by SciPy \cite{scipy} as a part of Sage Mathematics Software \cite{sage}.
If $\delta$ is too large the minimization algorithm 
fails to converge since the initial guess is too far away from 
the minimum, even in small networks.
The computational cost for solving the numerical problem
of finding $\{q_{ij}^*\}$ in our approach is large.
The time complexity grows as a stretched exponential of the network size, $t \sim \exp(an^{1/2})$\footnote{See Supplemental
Material at [URL {\em will be inserted by publisher}] for more details about the time
complexity of the algorithm.}.
In practice, this prevents the problem from being solvable in large networks. At variance, 
other related  problems \cite{Estrada2012, Rosvall2005} can be solved in polinomial time.
The average walk length is obtained by a Monte Carlo procedure using the set $\{q^*_{ij}\}$ 
for the transition probabilities. This quantity is analyzed as a function 
of the penalty and the shortest path length between the sources and the target.
%

\begin{figure}
\epsffile{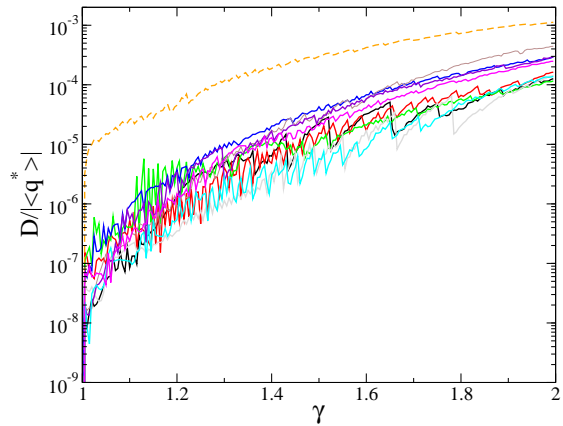}
\caption{\label{fig5}
(Color online) Each curve in this figure represents the ratio $D/\langle \vec{q^*}\rangle$ (see text) as a function of $\gamma$
for different targets $t$. One can see this ratio remains small in the whole range of $\gamma$
we explored.  The largest value of the ratio is $\simeq 0.001$ (orange dotted line).
The calculations were performed on a random network with $n=10$
and $\langle k\rangle = 3$.}
\end{figure}

The values of the probabilities $\{q_{ij}^*\}$ depend on the target vertex $t$
but are independent of the source vertex $s$ for each value of $\gamma$. 
We tested this analytically in the examples of
Fig. \ref{fig1}, and numerically on a small random network. 
Let us define the vector $\vec{q^*}$ whose components are the no-null values of $\{q_{ij}^*\}$.  
In Fig. \ref{fig5} we plot the ratio between the dispersion of $\vec{q^*}$ with respect to $s$, 
$D(\vec{q^*}) = 1/(n-2)\sum_{s\neq t} |\vec{q^*}(s)-\langle \vec{q^*}\rangle|^2$, 
and the norm, $\langle \vec{q^*}\rangle = 1/(n-1)\sum_{s\neq t} \vec{q^*}(s)$, 
as a function of $\gamma$ and for different targets $t$. 
It is shown that this ratio is much smaller than the unity confirming the independency of $\{q_{ij}^*\}$
with respect to the source.
%
%

Since information is an additive quantity, the relative amount of intrinsic information for the overall 
network can be defined,  given target $t$. It considers all the possible sources and hence the paths 
to the target $t$. Consequently  we have
\begin{equation}
\label{eq18}
R_{\gamma}(t) = \frac{\sum_{i\neq t} F_{\gamma}(i\to t)-S(i\to t)}{\sum_{i\neq t} F_{\gamma}(i\to t)},
\end{equation}
\noindent and then the overall optimal penalization $\gamma^*_t$ associated to the 
target $t$ can be obtained from this expression. For the sake of brevity let us omit the 
reference to $t$, so we will refer to $R_{\gamma}(t)$ by $R_{\gamma}$ and its 
optimal version $R^*(t)$ by $R^*$. 

\subsection{Random networks}

\begin{figure}
\epsffile{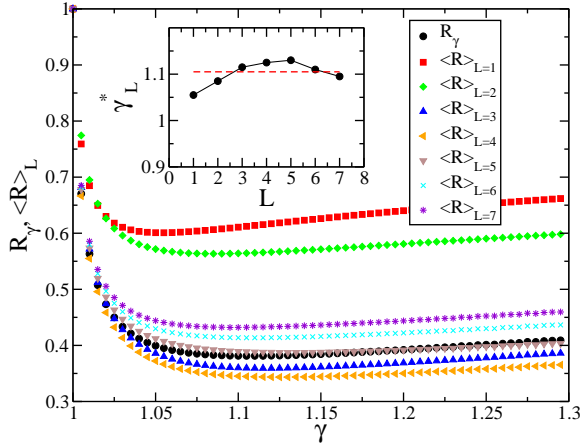}
\caption{\label{fig6}
(Color online) In black full circles the relative amount of intrinsic information $R_{\gamma}$
associated with a randomly chosen vertex $t$ [Eq. \ref{eq18}]. It shows
the typical behavior with a minimum at $\gamma^*_t \simeq 1.105$.
The other curves correspond to the relative amount of intrinsic information averaged over
all the sources at a fixed distance $L=1,2,...$ from $t$.
{\em Inset:} The minimum of $\avrg{R}_L$; $\gamma^*_L$,  depends on $L$, but is
the same order as $\gamma^*_t$ (red dotted line).
}
\end{figure}

Let us first analyze the case of a random network. All the calculations in this section 
were performed using a network of $N=100$ vertices with an average degree $\avrg{k}=3$ 
and using a target chosen at random. Care was taken to obtain a random network that consists
of only one connected component. Here we show results corresponding to a single 
realization of the target and the network, since similar results were obtained using different realizations.
Figure \ref{fig6} shows  $R_{\gamma}$ as function of $\gamma$ obtained for this network. 
From this curve we obtained the optimal overall penalization $\gamma^*_t\simeq 1.105$. 
In this figure we also show $\avrg{R}_L$, which is the relative information 
$R_s$ averaged over the sources
that are at a fixed distance $L$ from the target. We obtained from these curves the optimal 
penalization  $\gamma^*_L$;  the dependence of  $\gamma^*_L$ on $L$ is shown in the inset. 
We observe that  $\gamma^*_L$ varies with $L$ but in every case $\gamma^*_L$ is of the
same order of magnitude that $\gamma^*_t$. 
%
%

\begin{figure}
\epsffile{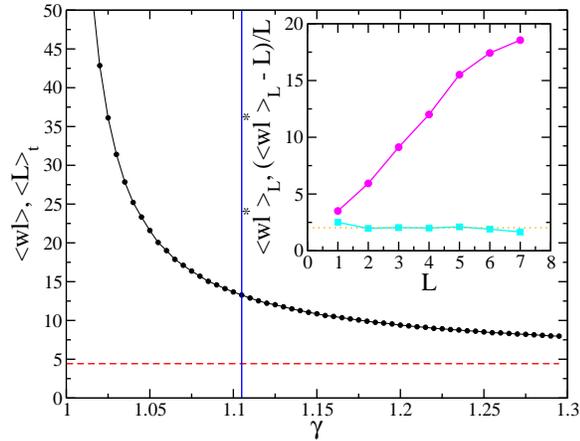}
\caption{\label{fig7}
(Color online) In black full circles the walk length $\avrg{wl}$ averaged over different sources
for a fixed target $t$ as a function of the penalization $\gamma$. The red dashed line
indicates average shortest path length $\avrg{L}_t$ to $t$.
The vertical blue full line indicates the optimal value of the
penalization $\gamma^*_t$. {\em Inset:} The magenta full circles correspond to the average walk length at
optimality $\avrg{wl^*}_L$ as a function of the distance to $t$. Cyan full squares represents the
relative difference $(\avrg{wl^*}_L - L)/L$ and the orange dotted line indicates the value $2$.
}
\end{figure}

\begin{figure}
\epsffile{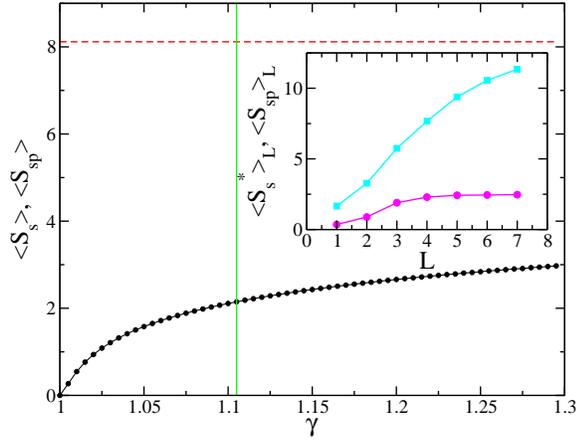}
\caption{\label{fig8}
(Color online) Walker's information $\avrg{S_s}$ averaged over all the sources
as function of the penalization $\gamma$ (black full circles). The horizontal red dashed line indicates
the average shortest path information $\avrg{S_{sp}}$. Vertical green full line indicates
the optimal penalization $\gamma^*_t$.
{\em Inset:} Average walker's information $\avrg{S^*_s}_L$
at optimal penalization $\gamma^*_t$ as function of the distance $L$ (magenta full circles) and
shortest path information  $\avrg{S_{sp}}_L$ (cyan full squares).
}
\end{figure}

In order to analyze the role of the penalty in the restricted walks toward the
target, we analyze typical walk lengths as function of $\gamma$. 
We first obtain the transition probabilities associated with the target $t$, for a given  $\gamma$,
and then using this probabilities we implement a Monte Carlo process to obtain a set of trajectories 
corresponding to random walkers which are biased to the target. The random walkers
start their journies at every possible source available in the network. When calculating the average of these 
trajectories  we obtain the average walk length $\avrg{wl}$. 
In Fig. \ref{fig7} we plot $\avrg{wl}$ corresponding to the same realization of the
network and target we used to obtain the results of Fig. \ref{fig6}.
When $\gamma$ approaches its minimal physical value $\gamma=1$ the walk is
fully random and the average walk length between all the possible sources and
the fixed target is $\avrg{wl} \simeq 260$, which is at least an order
of magnitude larger than the typical distance between vertices. Then as $\gamma$ is increased 
the average walk length  decreases drastically.  We denote by $\avrg{L}_t$ the average of all 
the shortest path that reach the target (note that in this calculation all sources are included).
%
%
At the optimal penalization  $\gamma^*_t$ the difference between the average walk length and the average
shortest paths $\avrg{L}_t$ is nearly two times $\avrg{L}_t$; that is, the average walk length is of the same
order of magnitude than the average of the shortest paths. We also computed the average walk length $\avrg{wl^*}_L$ 
at the optimal penalization $\gamma_t^*$
that correspond to averages in trajectories restricted to start
at sources that are at a distance $L$ from the target. 
We plotted this quantity as function $L$ in inset of Fig. \ref{fig7}.
The average walk length $\avrg{wl^*}_L$ increases linearly with the shortest path length
$L$, but the relative excess $(\avrg{wl^*}_L-L)/L$ is almost constant, taking a value close to $2$ 
(see inset of Fig. \ref{fig7}). 
%
%
In order to explore the dependency of the walker's information  $S$ on  
the penalty, we plot in Fig. \ref{fig8} the average, $\avrg{S_s}$, over all the sources 
 as a function  of $\gamma$. It can be seen that  $\avrg{S_s}$ grows 
with $\gamma$ but it is always much smaller than the averaged shortest path information
$\avrg{S_{sp}}$, in particular at the overall optimal penalization $\gamma^*_t$.
In the present approach the walker uses less information than in the 
shortest paths approach $S_{sp}$ but there is a price to pay for it; the path length 
to the target $\avrg{wl}$ is longer than the shortest path $L$. In addition,
according to the current approach the amount of information the walker learns on average
$\avrg{S_s^*}_L$ does not depend on $L$ (inset in Fig. \ref{fig8}) as long as $L$ is greater than $3$.
One can think that there is a distance horizon ($L=3$) that defines two regimes. At
short distances the walker can improve its information about the paths to the target
as the distance grows, whereas for targets far away the sources, the amount of information
cannot be improved. In other words, the searchability at short distance is favored \cite{Rosvall2005a,Rosvall2007}.

\subsection{Scale--free networks}
\begin{figure}
\epsffile{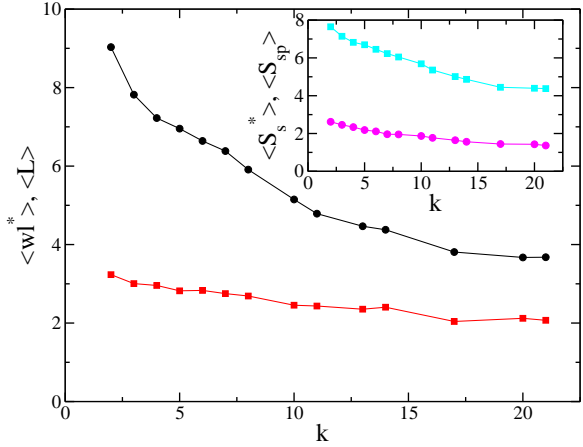}
\caption{\label{fig9}
(Color online) Average walk length $\avrg{wl^*}$ at optimal penalization
$\gamma^*_t$ (black full circles) and the average shortest
path length $\avrg{L}$ (red full squares) as a function of the target degree $k$.
Both quantities decrease as the degree increases.
{\em Inset:} Average walker's information $\avrg{S^*_s}$
at optimal penalization $\gamma^*_t$ (magenta full circles) and the shortest
path information $\avrg{S_{sp}}$ (cyan full squares) as a function of the target degree.
}
\end{figure}

\begin{figure}
\epsffile{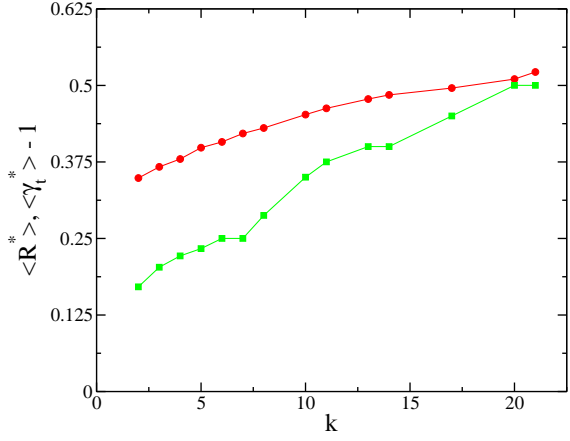}
\caption{\label{fig10}
(Color online) Average relative amount of intrinsic information  $\avrg{R^*}$
(red full circles) and average  of the overall optimal penalization $\avrg{\gamma^*_t}$ as a function of the
target degree $k$ (green full squares).
}
\end{figure}

Scale free networks are characterized by a power law distribution in the 
connectivity of the vertices, and even small networks shows the presence of highly 
connected vertices when compared to the mean value of their connectivity.
Therefore, although in our case the size of the network is small, a scale-free topology will allow us 
the study of how the different quantities are affected by the vertex's degree. 
We performed the calculations on a Barab\'asi-Albert network model\cite{Barabasi1999} with $N=100$ vertices 
and $\avrg{k}\simeq 4$. 
Figure \ref{fig9} shows that the walk length $\avrg{wl^*}$ and the 
shortest path length $\avrg{L}$ decrease with target's degree 
and the  decrease of the former is more pronounced. The hubs can be found easily 
having a walk length much closer to the shortest paths, whereas in poorly connected 
vertices significantly longer walks are required. According to the present approach the  
hubs are favored, regarding both the number of steps and the information 
that is needed, as far as the optimal condition is easily achievable.  
As in the case of the random network, in the scale-free
networks the walker's information $\avrg{S^*_s}$ is significantly smaller than the 
shortest path information $\avrg{S_{sp}}$ (see inset of Fig. \ref{fig9}). Both 
quantities decrease with the target's degree, confirming that highly connected 
vertices are easier to find. 
Since the shortest path information $\avrg{S_{sp}}$ varies more steeply than
the walker's information $\avrg{S^*_s}$ with the target's degree $k$, 
then it follows that $\avrg{S_{sp}}$ is more sensitive to
the topological details than $\avrg{S^*_s}$.
Finally, Fig. \ref{fig10} shows that the optimal penalization $\gamma_t^*$ and the relative
amount of intrinsic  information $R^*$ grows with the target's degree. This implies that it 
is more expensive to find the optimal walking pattern in highly connected targets.
As in scale-free networks, finite size effects may be very important, especially
for such small networks as used here. We repeated ten times the numerical calculations of Fig. \ref{fig9}
and \ref{fig10} ten times\footnote{See Supplemental Material at [URL {\em will be inserted by publisher}] for 
extentions of these calculations.}.
In order to check if the trend found is not due to the particular structure of the Barab\'asi-Albert
network model, we randomized the networks using the algorithm of Maslov-Sneppen \cite{Maslov2002}.
All the samples show the same tendency as the original calculation.

\begin{figure}
\epsffile{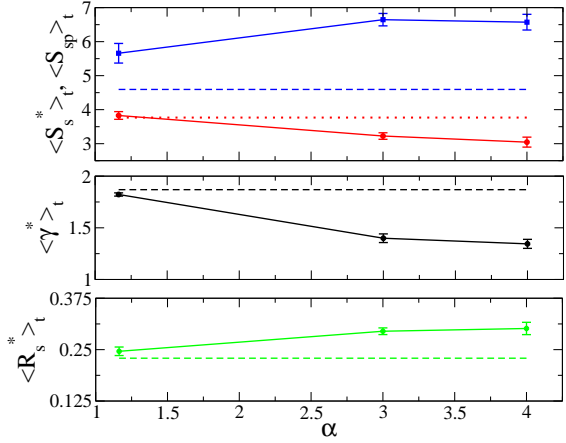}
\caption{\label{fig11}
(Color online) Comparison of different magnitudes of fat-tailed networks with degree distribution
$P(k)\sim (k_0+k)^{-\alpha}$ (symbols connected with full lines) against the star web (horizontal dotted and dashed lines) as
function of $\alpha$.
{\em Top:}
Optimal information $S^*(hub\to t)$ (red circles and dotted line) and shortest path information
$S_{sp}(hub\to t)$ (blue squares and dashed line) averaged over the targets.
{\em Center:}
Optimal penalty $\gamma^*$ averaged over the targets.
{\em Bottom:}
Optimal intrinsic information $R_s^*$ averaged over the targets.
}
\end{figure}

The scale-free networks should approach to the {\it star web} as the degree exponent increases. 
In order to test this hypothesis we generate scale-free tailed networks
\footnote{We follow the idea in Ref. \onlinecite{Rosvall2005a}.
A sample of the degree distribution $P(k)\sim (k_0+k)^{-\alpha}$ is obtained 
using the inverse transform sampling method. Then, a network is generated from this sample using the
Havel--Hakimi algorithm \cite{Havel1996} and randomized using the Maslov--Sneppen algorithm \cite{Maslov2002}.
Since the star web has $\avrg{k}=2$ we generate networks with connectivities near this value. 
This can be done in scale--free networks when $\alpha\simeq1.16$, but for larger values of $\alpha$ 
larger values of $\avrg{k}$ are required in order to obtain non fragmented networks. 
We choose $\avrg{k}(\alpha=1.16)=2,\avrg{k}(\alpha=3)=2.5$, and $\avrg{k}(\alpha=4)=2.75$ by means of properly 
setting the value of $k_0$.}$^,$
\footnote{See Supplemental Material at [URL {\em will be inserted by publisher}] for more details about the generated
networks.}
with varying exponent $-\alpha$.
In Fig. \ref{fig11} different quantities measured in the scale-free network are ploted as function 
of $\alpha$; as reference we also include the values corresponding to the {\it star web}. 
The magnitudes for the scale-free networks approach the values for the {\it star web} as $\alpha$ decreases.
In particular, the optimal information $\avrg{S_s^*}_t$ (shortest path information
$\avrg{S_{sp}}_t$) is smaller (greater) for the $\alpha$--dependent networks than for
the {\it star web}.
These two facts imply that the difference between the shortest path information $S_{sp}$
and the optimal information $S_s^*$ decreases as the hetereogeneity of the network grows.
In particular, the difference diminishes significatively only when $\alpha < 3$, which
is the range where the network heterogeneity is relevant as the variance $var(k)$ diverges for
infinite networks. The shortest path information for the $\alpha$-dependent networks is larger 
than that of the {\it star web} because inevitably, multiple steps along highly connected
nodes are required in the walk from the source to the target.

\section{Discussion and conclusions}
\label{conclu}
%

In this work we have introduced an approach for measuring the amount of information
used by a biased random walker that moves to a target. In this  framework, we extend the ideas 
of  Rosvall et al. \cite{Rosvall2005} because we consider not only the shortest paths
but all the possibles paths to the target.
Based in this approach we propose a penalization rule, which depends on one parameter and 
that bias a random walker to the target, provided the walker can use as little 
information as possible.
The basic idea was that each step that the walker takes is penalized;
hence, this leads to an overall penalty that tends to reduce the walk lengths.
Our approach is consistent since the two main quantities that determine the cost associated to the
task of targeted signaling are counterbalanced:
a shortening of the walk length through penalization implies an increase of the
information required, and vice-versa. 
At this point, it is important to stress that in this scheme the penalization operates globally, limiting the
overall available information, unlike other approaches \cite{Rosvall2005a} in which the information at each vertex is limited.
This has the advantage of overcoming the undesired effect of affecting mostly the highly
connected vertices, provided  the vertex's degrees are taken into account \cite{Rosvall2005a,Cajueiro2009}.
We also introduce the idea of  {\it intrinsic} information, in order to define an 
optimal penalization.
We have shown through some network models that in practice a small
amount of penalization is enough to drastically reduce the typical walk length,
and then a network can be efficiently navigated with restricted information.   
On the other hand, once the optimal penalization is reached it is highly
expensive to further reduce the typical walk length; in particular an infinite
penalization is required to restrict the path lengths to the shortest ones. 
The typical walk length in a random network was analyzed and compared to the corresponding
shortest path at the optimal penalization. It is found that the difference  
between these lengths grows linearly with the shortest path length. 
This is connected with a trade-off at which the amount of information does not increase with
the length of the shortest paths.
In addition, from the trend of $\avrg{S}_L$ a distance horizon can be identified which define 
a range of efficient searchability. It is worth stressing that the existence of an information horizon
has been previouly reported in the literature. In the context of targeted signaling by Trusina et al. \cite{Trusina2005}, and
in the complementary subject of efficient network diffusion by Sinatra et al. \cite{Sinatra2011} where an  
optimimal diffusion process is attainable with local limited information.         

The ideas introduced in this paper were applied to undirected and unweighted
networks.  However, in the study of  traffic on cities, directed and 
weighted  network are needed since streets have different capacities and directions. 
The extension of the formalism to include directed networks is
straightforward, but care must be taken to ensure that each vertex is accessible
from each other vertex; otherwise the analysis has to be restricted to each strongly 
connected components of the network. 
Also the penalty scheme may be generalized to be vertex dependent. 
In the case of the dual representation of network's cities
where vertices are streets and edges are road intersections\cite{Rosvall2005b}, the
traffic congestion on each street can be used to regulate the amount of
penalization in order to avoid a traffic jam.
Although the proposed formalism requires the solution of an optimization problem 
which has a large computational cost, the algorithm is specially suited 
for a parallel implementation since each target $t$ can be treated separately.

The authors thank Sergio Cannas, Daniel Stariolo and Pablo Serra for useful discussions. 
This work was partially supported by grants from CONICET (Argentina), 
Agencia C\'ordoba Ciencia (Argentina), SeCyT, Universidad Nacional de C\'ordoba (Argentina).

%
%

\end{document}